\documentclass[letterpaper]{article}

\usepackage{PRIMEarxiv}

\usepackage{graphicx}
\usepackage{array}
\usepackage{tabularx}
\usepackage{longtable}
\usepackage{xurl}
\usepackage{amsmath}
\usepackage{enumitem}

\usepackage[utf8]{inputenc} 
\usepackage[T1]{fontenc}    
\usepackage{hyperref}       
\usepackage{url}            
\usepackage{booktabs}       
\usepackage{amsfonts}       
\usepackage{nicefrac}       
\usepackage{microtype}      
\usepackage{fancyhdr}       
\graphicspath{{images/}}     

\title{A High-Resolution Synthetic EV Charging Dataset for Cold-Climate Distribution Grid Impact Analysis: Trondheim, Norway (2020--2030)
}

\author{
Hanieh Taraghi, Petr Musilek \\
  Department of Electrical and Computer Engineering \\
  University of Alberta \\
  Edmonton, Alberta, Canada\\
  \texttt{\{htaraghi@ualberta.ca, pmusilek@ualberta.ca\}}
}

\begin{document}
\maketitle

\begin{abstract}
This data article presents a high-resolution, long-term synthetic electric-vehicle (EV) charging dataset for Trondheim, Norway, spanning February 2020 to December 2030. Empirically grounded in 14 months of historical charging logs from December 2018 to January 2020, the dataset captures session-level behavioral patterns, including delivered energy, plug-in duration, connection schedules, user categorization (private vs. shared), seasonal variations, public-holiday effects, and daily ambient temperature dependencies. To model future electrification dynamics, the synthetic generation pipeline integrates historical session records, calendar and weather features from MET Norway, annual EV-adoption growth multipliers derived from Statistics Norway (SSB) registration trajectories, a daily session-count model, a Conditional Tabular Generative Adversarial Network (CTGAN), seasonal Kernel Density Estimation (KDE), and post-generation physical charger-power feasibility correction. Under a standardized 7.2~kW AC charging constraint, the resulting medium EV-adoption scenario dataset contains 76,993 hourly charging-activity records. The hourly profile is activity-based rather than a complete continuous hourly time series; hours with no allocated EV charging energy are not included. The records provide total hourly charging energy, equivalent average charging power, active session counts, private/shared user load breakdowns, ambient temperature features, and calendar indicators. The dataset provides a validated cold-climate benchmark for distribution-grid impact assessment, transformer-loading analysis, EV charging-demand forecasting, charger-capacity planning, energy-management optimization, and the development of data-driven smart-charging control strategies.
\end{abstract}


\begin{figure}[htbp]
    \centering
    \includegraphics[width=0.95\textwidth]{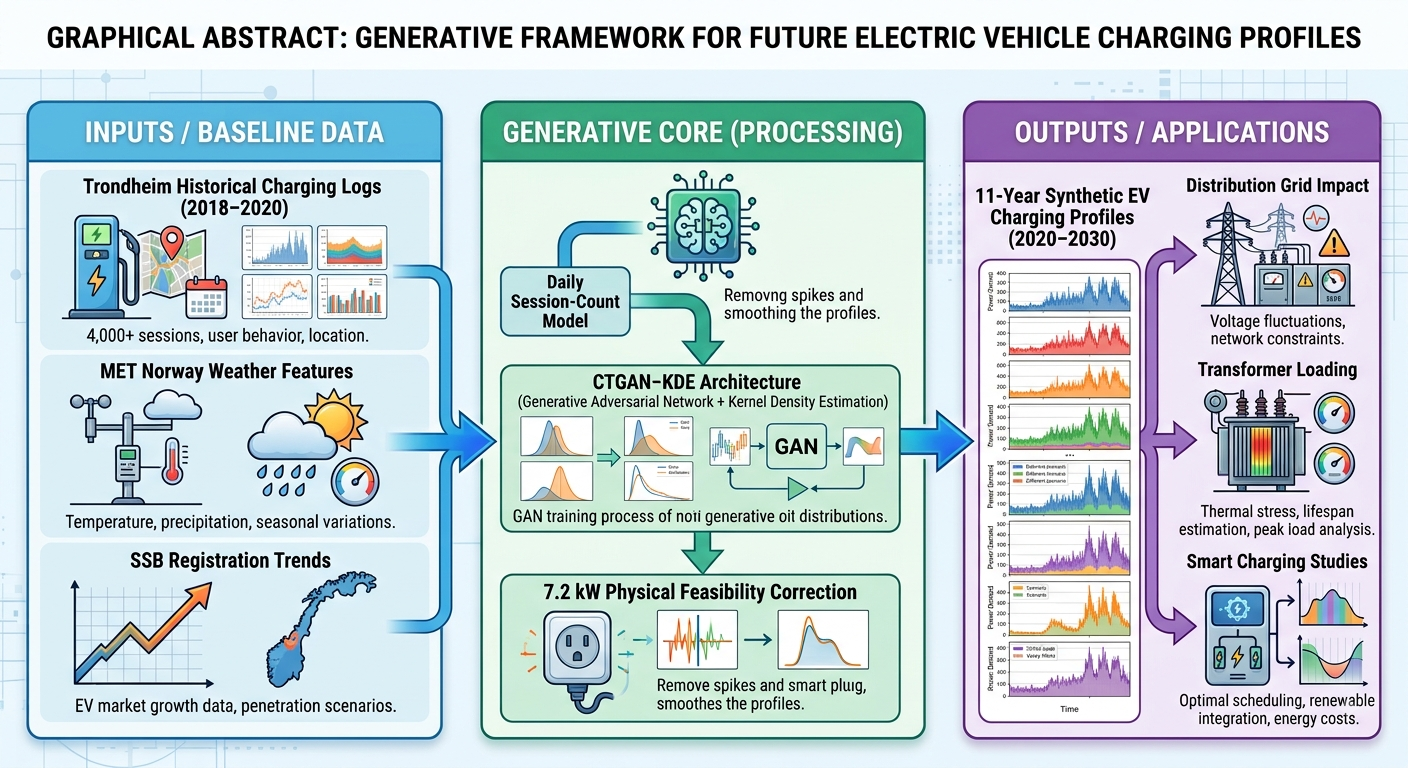}
\end{figure}

\begin{table}[h]
\centering
\caption{Specifications for the synthetic electric-vehicle charging dataset.}
\label{tab:specifications_table}
\resizebox{\textwidth}{!}{%
\begin{tabular}{ll}
\toprule
\textbf{Subject Area} & Energy and Power Systems, Artificial Intelligence \\
\midrule
\textbf{Specific Subject Area} & \begin{tabular}[t]{@{}l@{}}Electric Vehicle Charging Demand Modeling, Synthetic Data Generation,\\ Distribution Grid Impact Assessment, Cold-Climate Energy Systems\end{tabular} \\
\midrule
\textbf{Type of Data} & Table, Time-Series Data Profiles, Synthetic Behavioral Logs \\
\midrule
\textbf{Data Format} & Raw (.csv), Filtered (.csv), Aggregated (.csv) \\
\midrule
\textbf{Data Collection Methods} & \begin{tabular}[t]{@{}l@{}}Hybrid CTGAN--KDE generative architecture trained on historical Trondheim\\ charging logs (2018--2020), scaled via Statistics Norway (SSB) EV adoption\\ trajectories, enriched with MET Norway Frost API daily weather features, and\\ constrained by physical 7.2~kW AC charger power limits.\end{tabular} \\
\midrule
\textbf{Data Source Location} & \begin{tabular}[t]{@{}l@{}}\textbf{City/Region:} Trondheim, Tr{\o}ndelag \\ \textbf{Country:} Norway \\ \textbf{Coordinates:} 63.4305$^\circ$ N, 10.3951$^\circ$ E\end{tabular} \\
\midrule
\textbf{Data Accessibility} & Repository Name: Zenodo (Public Access) \\
\midrule
\textbf{Data Repository URL} & \url{https://doi.org/10.5281/zenodo.22132618} \\
\midrule
\textbf{Direct URL to Data} & \url{https://zenodo.org/records/22132618} \\
\midrule
\textbf{Instructions for Accessing Data} & Open access repository. All dataset files and supporting metadata are freely available. \\
\midrule
\textbf{Related Research Article} & N/A (Original Data Article) \\
\bottomrule
\end{tabular}%
}
\end{table}

\section*{Value of the Data}

\begin{itemize}
    \item \textbf{Long-Term Demand Horizons:} Provides a high-resolution, 11-year synthetic EV charging profile (2020--2030) calibrated to Trondheim, Norway, enabling multi-year distribution grid planning and long-term infrastructure investment studies.
    \item \textbf{Behavioral \& Contextual Fidelity:} Captures multidimensional driver behavior---including temporal connection patterns, user categories, seasonal dynamics, and holiday effects---derived from empirical charging logs via a hybrid CTGAN--KDE architecture.
    \item \textbf{Cold-Climate Parameterization:} Integrates daily temperature dependencies and cold-weather scaling factors, providing a specialized benchmark dataset for cold-climate EV load forecasting and thermal impact analyses.
    \item \textbf{Multi-Scenario Penetration Modeling:} Incorporates parametrized EV growth trajectories ($G_y$) to model the impact of escalating fleet penetration on aggregated hourly power demand, peak-to-average load ratios, and station utilization rates.
    \item \textbf{Physical Feasibility Bounding:} Enforces individual hardware limits ($P_{\mathrm{charger}} = 7.2~\text{kW}$) post-generation, ensuring that non-linear charging dynamics remain physically realistic for downstream power flow modeling.
    \item \textbf{Multi-Scale Application Readily:} Structured at both session-level and hourly aggregate formats to directly interface with distribution system power flow models, transformer degradation assessments, smart-charging control algorithms, and reinforcement learning environments.
\end{itemize}

\section*{Background}
Electric-vehicle (EV) charging datasets are a core input for analyzing charging demand, charging-session behavior, station utilization, and the interaction between transport electrification and distribution networks. High-resolution charging records typically contain plug-in time, plug-out time, charged energy, session duration, and user- or station-level identifiers, enabling detailed evaluations of temporal demand patterns and user heterogeneity \cite{sorensen2024norwaydataset,gholizadeh2024daily}. For example, S{\o}rensen \textit{et al.} released a Norwegian residential EV charging dataset containing more than 35{,}000 charging sessions across 12 locations \cite{sorensen2024norwaydataset}. Similarly, Mystakidis \textit{et al.} analyzed the Risvollan residential complex in Trondheim using session data alongside local traffic, weather, and user-behavior variables, demonstrating the value of contextual data for behavior-aware EV charging-demand forecasting \cite{mystakidis2025forecasting}.
However, observed EV charging datasets are frequently restricted to specific geographic contexts, infrastructure configurations, and observation windows, limiting their direct transferability to long-term planning studies under expanding EV adoption \cite{li2024synthesis,alaraj2025review}. This constraint is particularly critical for forecasting applications, where model performance depends heavily on data availability, aggregation scale, and the representativeness of historical records \cite{ostermann2024probabilistic,alaraj2025review}.

These limitations have motivated growing interest in synthetic and AI-augmented EV charging datasets. Li \textit{et al.} proposed a real-world data-driven synthesis framework trained and validated on approximately 1.65 million charging events, showing that synthetic samples preserve the distributional structure of observed charging behavior and can be conditionally generated for future scenarios \cite{li2024synthesis}. Gholizadeh and Musilek generated a 29{,}600-day synthetic daily charging dataset from Caltech charging logs using conditional tabular generative adversarial networks (CTGAN) and kernel density estimation (KDE), preserving connection times, charging durations, and energy demand under station-capacity constraints \cite{gholizadeh2024daily}. More recently, Liu \textit{et al.} introduced MP-EVData, combining station-level charging profiles across 10 prototype configurations with AI-augmented synthetic profiles, demonstrating the utility of synthetic data for data-intensive forecasting, smart charging, and infrastructure planning while maintaining daily, weekly, and annual temporal structures \cite{liu2026mpevdata}. Collectively, these studies show that synthetic EV charging datasets are most effective when they preserve session-level statistics and temporal regularities while scaling to future demand levels.

The need for high-fidelity synthetic data is particularly pronounced in power-system studies. Unmanaged EV charging can exacerbate peak demand, voltage deviations, line and transformer loading, and distribution network congestion; thus, accurate charging-load forecasting is essential for grid operations, charging management, and infrastructure planning \cite{ahmed2026review,ostermann2024probabilistic,mao2026integrated,dong2025deepexpert,he2026multitask}. In this context, realistic hourly load profiles directly support transformer-loading assessments, feeder-congestion analyses, flexible-charging design, and probabilistic planning, while complementing session-level records that capture individual charging behavior \cite{ahmed2026review,ostermann2024probabilistic,mao2026integrated}. Recent forecasting literature has consequently shifted toward probabilistic, multi-scale, and multi-task formulations to capture uncertainty, temporal dynamics, station heterogeneity, and aggregate energy consumption \cite{ostermann2024probabilistic,dong2025deepexpert,he2026multitask}.

Cold-climate and seasonal regions require specialized modeling because ambient temperature significantly influences vehicle energy consumption and charging demand. Jack \textit{et al.} developed a degree-day model from multi-year records of over 1{,}000 vehicles, finding that EV charging demand varies seasonally by up to 16\% across regions and increases winter monthly consumption by up to 30\% \cite{jack2025seasonal}. For infrastructure planning, Dimatulac \textit{et al.} demonstrated that winter conditions exacerbate capacity bottlenecks for long-haul EV operations in Ontario \cite{dimatulac2026ontario}. Environmental and calendar dependencies are also evident in broader public-charging studies: Feng \textit{et al.} showed that public charger utilization varies significantly across regions in Great Britain depending on environmental and temporal factors \cite{feng2026greatbritain}, while Liu \textit{et al.} reported distinct daily, weekly, and annual patterns across station prototypes \cite{liu2026mpevdata}. More broadly, machine-learning forecasting literature identifies weather, weekends, holidays, traffic, and calendar variables as key explanatory features for charging-demand estimation \cite{alaraj2025review,mystakidis2025forecasting}.

For a cold-climate urban context such as Trondheim, the literature supports a locally grounded synthetic dataset design that integrates observed charging behavior with Norwegian calendar structures, temperature-dependent demand adjustments, and EV-adoption scaling. Evidence from Norway indicates that driver behavior is highly heterogeneous: Zatsarnaja \textit{et al.} identified three distinct driver classes with significant differences in charging frequency, location preference, and target state-of-charge \cite{zatsarnaja2025norway}. Synthesizing these insights, a Trondheim-oriented synthetic dataset should avoid simple mechanical extrapolation of historical sessions. Instead, it must preserve local session statistics, encode behavioral, calendar, and weather variations, enforce physical charger-power limits, and aggregate sessions into hourly load profiles suitable for grid-impact and smart-charging analyses \cite{zatsarnaja2025norway,mystakidis2025forecasting,li2024synthesis,jack2025seasonal,ahmed2026review}.

\section*{Data Description}

\subsection*{Dataset Content}

The primary output of this study is a validated synthetic hourly electric vehicle (EV) charging-load dataset for Trondheim, Norway. The core dataset, designated as \texttt{synthetic hourly aggregated medium 7.2kW}, models a medium EV-adoption scenario under a 7.2~kW AC charger power limit. This aggregate profile is derived from the session-level dataset \texttt{synthetic sessions 2020-2030 medium 7.2kW corrected}, which comprises individual synthetic charging sessions following physical feasibility correction.

The aggregated dataset contains 76,993 hourly records. The aggregated dataset contains 76,993 hourly records. The aggregated profile is activity-based rather than a complete continuous hourly time series; hours with no allocated EV charging energy are not included. Each record corresponds to a single hourly interval and includes:

\begin{itemize}
\item \textbf{Temporal and Calendar Metrics:} Interval start/end timestamps, date, year, month, day, hour, weekday number, weekday name, weekend status, and season.
\item \textbf{Charging Load Metrics:} Total hourly energy demand, mean hourly power, total active session count, private charging energy, shared charging energy, private active session count, and shared active session count.
\item \textbf{Environmental and Contextual Variables:} Daily mean temperature for Trondheim, EV adoption scenario identifier, charger case classification, and physical charger power limit.
\end{itemize}

The temporal scope covers synthetic EV charging demand from February 2020 through December 2030. A minor subset of hourly records extends into early January 2031 to account for multi-day sessions initiated on December 31, 2030, thereby ensuring strict energy conservation between the session-level and hourly aggregated datasets.

The corrected session-level dataset comprises 256,826 synthetic charging sessions, yielding a cumulative energy demand of approximately 3.26~GWh across the 11-year horizon. The mean hourly total charging energy is 42.33~kWh, with a peak hourly aggregated demand of 1559.87~kWh. This maximum value reflects regional future EV demand rather than a transformer-limited or site-constrained capacity. If feeder, transformer, or station-level capacity constraints are required, these should be applied as secondary physical bounds in downstream grid-impact analyses.

Validation confirmed that all corrected sessions satisfy the 7.2~kW power constraint and that total energy is identically conserved between the session-level logs and hourly aggregate profiles. Consequently, the dataset is directly applicable to EV charging-demand forecasting, distribution-grid impact studies, transformer-loading assessments, and the development of data-driven energy-management or reinforcement-learning models.

\section*{Experimental Design, Materials, and Methods}

The synthetic dataset was generated using a structured workflow designed to preserve observed Trondheim EV charging behavior while extending the profile through 2030. Initially, the historical session logs were cleaned by removing records with missing plug-in or plug-out timestamps, missing durations, negative delivered energy, non-positive plug-in durations, inconsistent timestamps, and duplicate session identifiers. Long-duration sessions and sessions exhibiting high average charging power were retained but flagged, as these records may represent authentic charging behavior rather than data anomalies.

Calendar features were constructed for both baseline and future synthetic periods. Each day was categorized by year, month, day, day of year, week of year, weekday, weekend status, and season (Winter: December--February; Spring: March--May; Summer: June--August; Autumn: September--November). Norwegian national public holidays and special calendar periods, including Christmas/New Year, Easter, summer holidays, and bridge days, were incorporated to capture calendar-dependent demand shifts.

Daily weather features were integrated using Trondheim temperature observations retrieved via the MET Norway Frost API \cite{met_frost}. Frost provides access to the MET Norway archive of historical weather and climate observations. For the baseline observation period, daily temperature variables were obtained from weather stations representing Trondheim using daily aggregated air-temperature observations. Extracted metrics included daily mean, minimum, and maximum temperatures, heating-degree days (HDD), cold-day indicators, very-cold-day indicators, and a temperature-based charging-demand multiplier. For future years, historical Trondheim weather sequences were resampled to simulate plausible daily weather trajectories. The temperature multiplier was subsequently applied to scale expected charging demand during cold conditions.

EV adoption growth was calculated using registered electric passenger-car stock data from Statistics Norway (SSB) \cite{ssb_07849,ssb_registered_vehicles}. The primary data source was SSB StatBank Table 07849, which reports registered vehicles by year, region, transport type, and fuel type \cite{ssb_07849}. In this study, the vehicle category was filtered to private passenger cars, fuel type to electricity, and municipality to Trondheim. If complete Trondheim-level data were unavailable, Tr{\o}ndelag county data served as a fallback, while national figures for Norway were retained for comparison. As SSB reports private-car stock as of 31 December, these figures were treated as end-of-year registered electric car stock \cite{ssb_07849}. Yearly EV stock is denoted as $\text{EV}_y$, where $y$ represents the target year.

Two EV-growth indicators were calculated. First, the annual EV growth rate $g_y$ was calculated as:
\begin{equation}
g_y = \frac{\text{EV}_y - \text{EV}_{y-1}}{\text{EV}_{y-1}} \times 100,
\end{equation}
where $g_y$ represents the percentage growth rate from year $y-1$ to year $y$, $\text{EV}_y$ is the number of registered electric private cars in year $y$, and $\text{EV}_{y-1}$ is the registered stock in the preceding year. Second, an EV-growth multiplier $G_y$ was defined as:
\begin{equation}
G_y = \frac{\text{EV}_y}{\text{EV}_{2019}}.
\end{equation}
The year 2019 served as the baseline ($G_{2019}=1.00$), representing the primary complete historical year in the observed dataset. Values exceeding 1.00 indicate relative growth in registered EV stock compared to 2019. This multiplier scaled the expected daily synthetic session count rather than directly altering individual session energy levels. For years beyond available SSB observations, conservative, medium, and high EV-adoption trajectories were projected to account for adoption uncertainty through 2030.

A daily session-count model estimated the volume of charging sessions for each synthetic day. Baseline daily charging intensity was parameterized from 2019 historical observations as a function of month and weekday. The expected daily session count for future dates was then scaled by the EV-growth multiplier $G_y$, temperature multipliers, and holiday factors. To address overdispersion in the historical daily session counts, a Negative Binomial distribution was employed for stochastic sampling.

Synthetic session-level attributes were generated via a hybrid CTGAN--KDE approach. A conditional CTGAN model was trained on cleaned historical sessions enriched with calendar, seasonal, holiday, and weather features. The model learned the joint distribution across delivered energy, plug-in duration, arrival time, departure time, user type, garage identifier, month, weekday, season, holiday indicators, and temperature metrics. To enforce season-specific dynamics, separate multivariate kernel density estimation (KDE) models were fitted for each season using delivered energy, plug-in duration, arrival time, departure time, and average charging power. During generation, CTGAN outputs were blended with KDE samples corresponding to the target season to maintain both multivariate relationships and empirical seasonal tails.

Physical feasibility constraints were enforced post-generation. Each synthetic session was evaluated against the charger-power limit:
\begin{equation}
E \leq P_{\mathrm{charger}} D,
\end{equation}
where $E$ represents delivered energy (kWh), $P_{\mathrm{charger}}$ denotes charger power (kW), and $D$ is plug-in duration (hours). Under the primary 7.2~kW AC charging case, maximum deliverable energy was constrained to $7.2D$. If a generated session exceeded this limit, delivered energy was capped at the physically feasible maximum, and average charging power was updated accordingly. Sessions were adjusted rather than removed to maintain session-count structures and adoption assumptions. Due to changes in SSB municipality allocation rules for leased vehicles starting in statistical year 2025, 2024--2025 variances at the Trondheim and Tr{\o}ndelag levels were interpreted cautiously, with national values serving as reference \cite{ssb_registered_vehicles}.

Corrected session logs were converted into hourly aggregate load profiles. Delivered energy per session was distributed across hourly intervals between plug-in and plug-out timestamps while respecting the 7.2~kW AC power limit. Partial boundary hours were weighted by the active fraction of the hour. Aggregate hourly profiles were then constructed by summing across individual sessions to derive total hourly energy demand, mean hourly power, active session counts, and load breakdowns for private and shared charging categories.

Dataset validation comprised statistical, temporal, seasonal, and physical-consistency checks. Descriptive statistics and correlation heatmaps verified distribution alignment between historical and synthetic sessions. Energy conservation between session logs and aggregate hourly profiles was confirmed, and all corrected sessions satisfied the 7.2~kW AC power upper bound. Consequently, the validated dataset is suitable for EV charging-demand modeling, distribution-grid impact studies, transformer-loading assessments, and reinforcement-learning control applications.

\begin{table}[htbp]
\centering
\caption{Descriptive statistics comparison between observed Trondheim charging sessions and generated synthetic charging sessions.}
\label{tab:descriptive_statistics}
\resizebox{\textwidth}{!}{%
\begin{tabular}{lrrrrrr}
\toprule
& \multicolumn{3}{c}{\textbf{Observed Data}} & \multicolumn{3}{c}{\textbf{Generated Synthetic Data}} \\
\cmidrule(lr){2-4} \cmidrule(lr){5-7}
\textbf{Variable} & 
\textbf{Mean} & 
\textbf{Std.} & 
\textbf{Median} & 
\textbf{Mean} & 
\textbf{Std.} & 
\textbf{Median} \\
\midrule
Delivered energy (kWh) & 12.73 & 11.79 & 9.06 & 12.69 & 12.44 & 9.43 \\
Plug-in duration (h) & 11.50 & 14.15 & 10.03 & 11.69 & 13.07 & 9.47 \\
Arrival time (h) & 16.82 & 4.37 & 17.07 & 16.25 & 4.73 & 16.63 \\
Departure time (h) & 13.05 & 5.25 & 12.45 & 13.14 & 5.39 & 12.81 \\
Average charging power (kW) & 2.19 & 1.91 & 1.52 & 2.46 & 2.46 & 1.39 \\
\bottomrule
\end{tabular}%
}
\end{table}

Table~\ref{tab:descriptive_statistics} compares the primary descriptive statistics of the observed Trondheim charging sessions against the synthetic dataset. Overall, the synthetic dataset demonstrates close alignment with historical statistics, confirming that central behavioral characteristics are preserved.

For delivered energy, the observed mean is 12.73~kWh compared to a synthetic mean of 12.69~kWh. Medians show similar agreement (9.06~kWh observed vs. 9.43~kWh synthetic). The synthetic standard deviation is slightly higher (12.44~kWh vs. 11.79~kWh), reflecting extended variance introduced by future adoption scaling, weather integration, and stochastic sampling.

Plug-in durations exhibit close statistical agreement, with mean durations of 11.50~h (observed) and 11.69~h (synthetic), and medians of 10.03~h and 9.47~h, respectively. These metrics confirm that long-duration connection patterns are maintained.

Arrival and departure dynamics are similarly preserved. Mean arrival times are 16.82~h (observed) and 16.25~h (synthetic), aligning with late-afternoon return-home plug-in events. Mean departure times are 13.05~h (observed) and 13.14~h (synthetic).

Average charging power shows a modest increase in the synthetic mean (2.46~kW vs. 2.19~kW observed) and standard deviation (2.46~kW vs. 1.91~kW observed). This shift stems from the physical feasibility correction under a 7.2~kW AC charger cap, allowing higher average power for energy-intensive sessions while maintaining median values close to baseline (1.39~kW synthetic vs. 1.52~kW observed).

Overall, the statistical comparison confirms that the synthetic generation framework captures the central behavioral distributions of the baseline dataset while generalizing patterns to support future EV adoption scenarios.

\begin{figure}[htbp]
    \centering
    \includegraphics[width=0.95\textwidth]{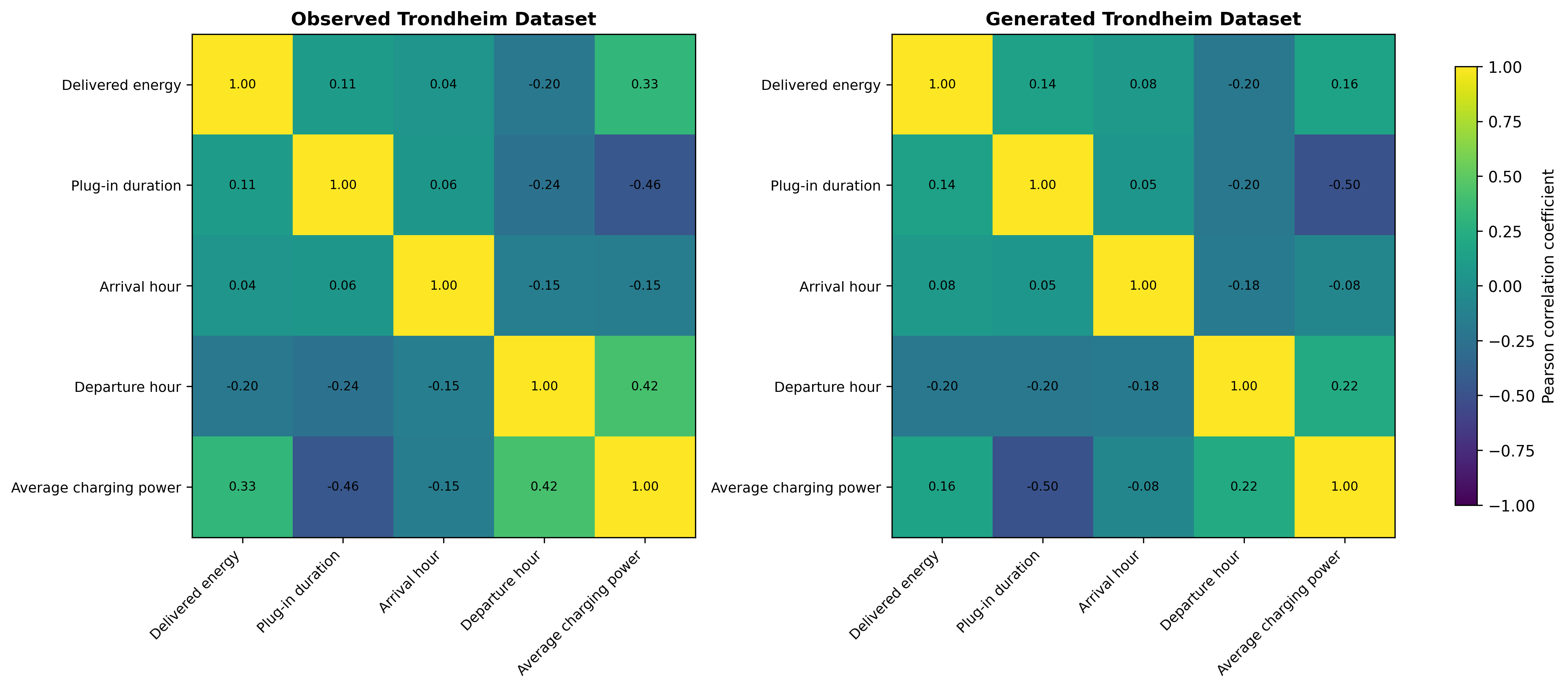}
    \caption{Correlation heatmap comparison between observed Trondheim charging sessions and generated synthetic charging sessions.}
    \label{fig:correlation_heatmap}
\end{figure}

Figure~\ref{fig:correlation_heatmap} compares the Pearson correlation matrices of the observed and synthetic charging sessions. The synthetic dataset reproduces the primary inter-variable dependencies of the baseline data.

In both datasets, delivered energy exhibits a weak positive correlation with plug-in duration ($r = 0.11$ observed vs. $r = 0.14$ synthetic), as longer plug-in durations allow for higher energy transfer despite extended idle connection periods. Delivered energy remains negatively correlated with departure hour ($r = -0.20$ across both datasets), reflecting higher energy delivery during overnight charging sessions ending early in the day.

A strong negative correlation exists between plug-in duration and average charging power ($r = -0.46$ observed vs. $r = -0.50$ synthetic), capturing the physical behavior wherein long-duration connections average lower power demand compared to short top-up sessions.

The positive correlation between departure hour and average power is preserved, though attenuated in the synthetic data ($r = 0.42$ observed vs. $r = 0.22$ synthetic). This attenuation is an expected outcome of stochastic sampling, weather adjustments, and 7.2~kW charger-power capping over the extended simulation horizon.

Correlations involving arrival hour remain low across both datasets ($r = -0.15$ observed vs. $r = -0.18$ synthetic for arrival vs. departure hour), indicating that arrival time primarily dictates temporal session placement rather than directly determining energy or duration magnitude.

\begin{figure}[htbp]
    \centering
    \includegraphics[width=0.95\textwidth]{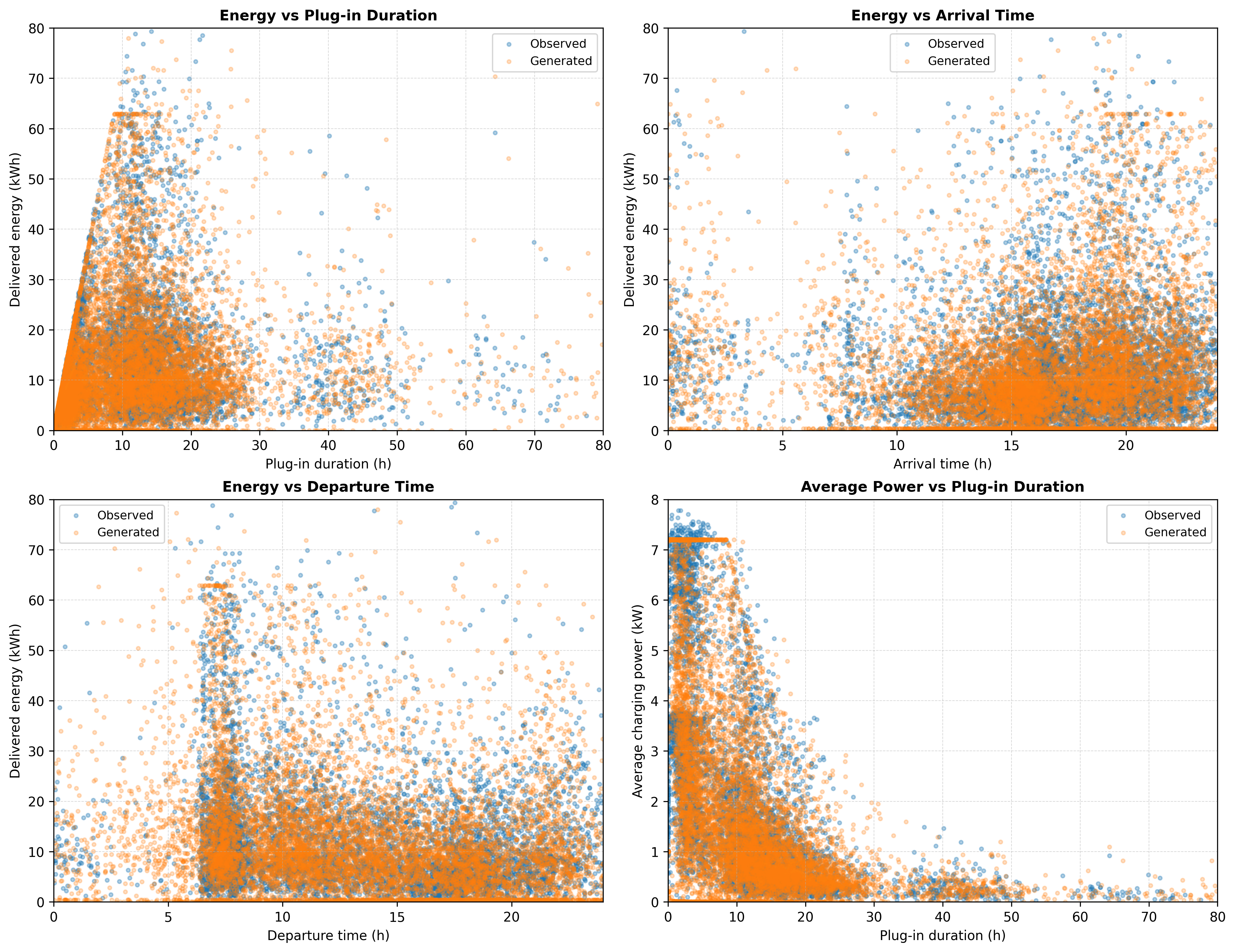}
    \caption{Scatter plot comparison between observed Trondheim charging sessions and generated synthetic charging sessions.}
    \label{fig:scatter_comparison}
\end{figure}

Figure~\ref{fig:scatter_comparison} presents pairwise scatter distributions for historical and synthetic sessions, illustrating consistent behavioral envelopes.

In the energy versus plug-in duration subplots, both datasets display a characteristic triangular distribution where energy demand scales with duration for short-to-medium sessions before dispersing at longer durations due to idle plug-in time. All synthetic sessions remain bounded by the 7.2~kW physical charger constraint.

The energy versus arrival time plots confirm peak arrival density between 14:00 and 22:00, characteristic of residential plug-in habits. Similarly, energy versus departure time displays concentrated morning departures across comparable energy ranges.

The average power versus plug-in duration profile demonstrates a consistent inverse relationship across both datasets, confirming that short sessions exhibit higher average power while long sessions remain dominated by low average power demand, bounded strictly by $P_{\mathrm{charger}} = 7.2~\text{kW}$.

\begin{figure}[htbp]
    \centering
    \includegraphics[width=0.95\textwidth]{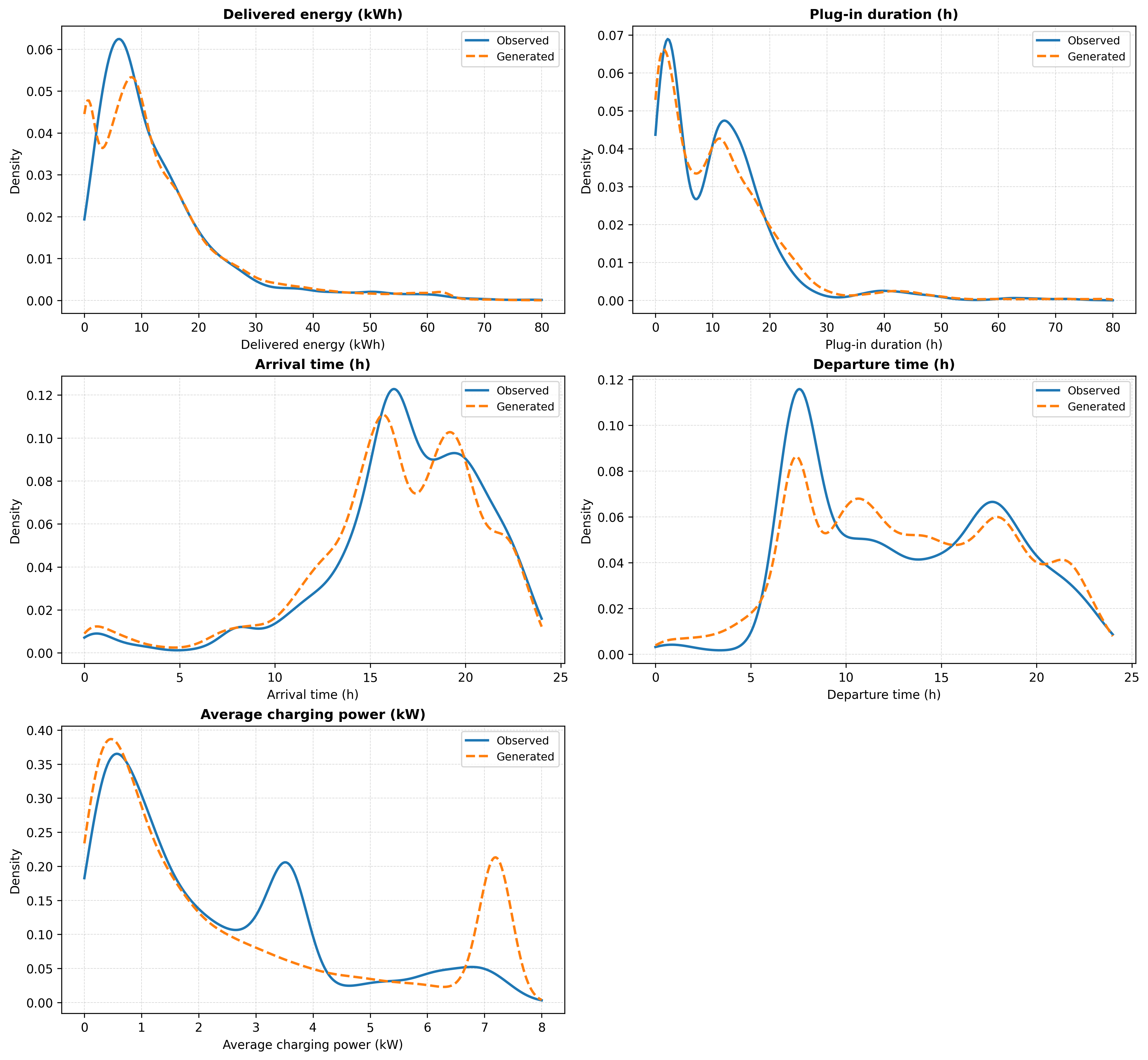}
    \caption{Density plot comparison between observed Trondheim charging sessions and generated synthetic charging sessions.}
    \label{fig:density_comparison}
\end{figure}

Figure~\ref{fig:density_comparison} compares marginal probability densities between historical and synthetic datasets.

Delivered energy exhibits a right-skewed profile concentrated below 20~kWh with a smooth upper tail. Plug-in duration shows a bimodal distribution reflecting short top-up sessions alongside extended parking events.

Arrival and departure densities maintain peak timing characteristics: arrival hours peak in the late afternoon (16:00--18:00), while departure hours show strong morning density. 

Average charging power exhibits high density at low values ($< 2~\text{kW}$), with a secondary density concentration at the 7.2~kW limit resulting from the physical energy-capping feasibility correction.

\section*{Limitations}

Several limitations should be considered when applying or extending this dataset:

\begin{enumerate}[label=\roman*.]
    \item \textbf{Baseline Historical Span:} The generative models rely on historical Trondheim charging records collected between December 2018 and January 2020. While the generative pipeline projects scenarios through 2030, long-term shifts in battery chemistry/capacity, vehicle-to-grid (V2G) adoption, real-time tariff structures, and driver routing behavior are not explicitly modeled.
    \item \textbf{Adoption Dynamics:} EV stock expansion is modeled using annual municipal growth multipliers ($G_y$) that scale overall daily session frequencies. This macro-level formulation does not dynamically simulate micro-level agent decisions, public-versus-private charger infrastructure expansion, or spatial redistribution across urban zones.
    \item \textbf{Grid Bounds:} Unconstrained hourly loads represent unmitigated aggregated charging demand. Physical feasibility corrections enforce a 7.2~kW per-charger hardware limit, but secondary distribution network constraints (e.g., transformer kVA capacity, feeder current limits, or phase unbalance) are omitted and must be enforced in downstream power-flow or optimization studies.
    \item \textbf{Exogenous Weather Features:} Environmental influences are captured via daily mean temperature and cold-weather demand scaling factors. Additional meteorological variables, such as snowfall depth, precipitation, road surface friction, and extreme wind conditions, were not explicitly modeled.
    \item \textbf{Anonymization and Behavioral Abstraction:} Synthetic session identifiers, garage keys, and individual profiles preserve population-level joint probability distributions rather than exact physical driver schedules. Individual synthetic trajectories should be treated as representative behavioral samples rather than traceable end-user profiles.
    \item \textbf{Geographic Generalizability:} Parameterization is calibrated against climate and EV adoption profiles specific to Trondheim, Norway. Direct application to regions with distinct climate regimes, lower baseline EV penetration, or different grid architectures requires re-calibration against local baseline data.
\end{enumerate}

\section*{Ethics Statement}

This study utilized secondary, fully anonymized observational data and synthetic data generation techniques; it involved no active human-subject experimentation, animal testing, or collection of clinical/personally identifiable information (PII). All synthetic records generate representative behavioral patterns rather than individual user profiles. Original baseline datasets were processed in accordance with national data privacy standards, ensuring location-sensitive parameters were generalized or removed prior to model training and public repository release.

\bibliographystyle{plain}
\bibliography{references}

\end{document}